\documentclass[prb,onecolumn,showpacs,preprintnumbers,amsmath,amssymb,assume]{revtex4}
\usepackage{graphicx}				
\usepackage{dcolumn}			 
\usepackage{bm}						  
\usepackage{cases}
\usepackage{color}
\newcommand{\Bs}{&&\hspace{-3.2mm}} 

\begin{document}
	
	\title{Triple singularities of elastic wave propagation in anisotropic media}
	
	\author{Vladimir Grechka$^{1}$}
	\affiliation{$^{1}$Borehole Seismic, LLC}
	\date{\today}
	
\begin{abstract}

A typical singularity of elastic wave propagation, often termed a shear-wave singularity, takes place when the Christoffel equation has a double root or, equivalently, two out of three slowness or phase-velocity sheets share a common point. We examine triple singularities, corresponding to triple degeneracies of the Christoffel equation, and establish their two notable properties: (i)~if multiple triple singularities are present, the phase velocities along all of them are exactly equal, and (ii)~a triple singularity maps onto a \emph{finite-size planar} patch shared by the group-velocity surfaces of the P-, S$_1$-, and  S$_2$-waves. There are no other known mechanisms that create finite-size planar areas on group-velocity surfaces in homogeneous anisotropic media.

\end{abstract}
\pacs{81.05.Xj, 91.30.-f}   
\maketitle

\section{Introduction} \label{sec:intro}

Singularities of seismic wave propagation, defined as the wavefront normal directions along which the Christoffel equation has a double root, constitute a subject of quite an extensive literature\cite{Fedorov1968, AlshitsLothe1979a, AlshitsLothe1979, Crampin1981review, CrampinYedlin1981, Musgrave1981, Alshitsetal1985, Helbig1994, SchoenbergHelbig1997, Shuvalov1998, BoulangerHayes1998, Vavrycuk2001, Vavrycuk2003, Vavrycuk2005a, Vavrycuk2005weak, Alshits2003, Goldin2013, Grechka2015GofR, Grechka2017R2G, Ivanov2019}. The interest to singularities can be explained, at least partly, by their ubiquity~| they are found in all natural anisotropic elastic solids.

The so-called three-fold or triple singularities, where the Christoffel equation possesses a triple root, on the other hand, are mentioned in a handful of papers\cite{AlshitsLothe1979a, Musgrave1981, Alshitsetal1985, Alshits2003} only and not well known in both acoustical and geophysical communities. The reason for that, presumably, is the instability of triple singularities with respect to a small arbitrary perturbation of the stiffness tensor\cite{Alshitsetal1985}. Because the lack of their stability should not disqualify them from theoretical investigation, we present a study of their features. 

\section{Theory}

The classic Christoffel\cite{Christoffel1877, Fedorov1968, Musgrave1970} equation
\begin{equation} \label{eq:ani101-40}
\bm{\Gamma}(\bm{n}) \bm{\cdot U} = V^2 \, \bm{U} \, 
\end{equation}
describes propagation of plane body waves in homogeneous anisotropic media. Here
\begin{equation} \label{eq:ani101-41}
\bm{\Gamma}(\bm{n}) \equiv  \bm{n \cdot c \cdot n} \,
\end{equation}
is the $3 \times 3$ Christoffel tensor or matrix, $\bm{c}$ is the $3 \times 3 \times 3 \times 3$ density-normalized stiffness tensor, $\bm{n}$ is the unit normal to a plane wavefront, $V$ is the phase velocity, and $\bm{U}$ is the unit polarization vector. Mathematically, equation~\ref{eq:ani101-40} is the standard eigenvalue-eigenvector problem for the $3 \, \times \, 3$ symmetric, positive-definite matrix  $\bm{\Gamma}(\bm{n})$; the eigenvalues of $\bm{\Gamma}(\bm{n})$ are the squared phase velocities $V^2 > 0$ of the three isonormal plane waves, termed the P-, fast shear S$_1$-, and slow shear S$_2$-waves; the eigenvectors of $\bm{\Gamma}(\bm{n})$ are the unit polarization vectors $\bm{U}$ of these waves. 

Triple singularities are defined as wavefront normal directions $\bm{n} = \bm{n}^t$ along which  equation~\ref{eq:ani101-40} has three coinciding eigenvalues,
\begin{equation} \label{eq:sws-28}
V_\mathrm{P}^2(\bm{n}^t) = V_\mathrm{S1}^2(\bm{n}^t) = V_\mathrm{S2}^2(\bm{n}^t) \equiv V_t^2 \, .
\end{equation}
As a consequence of equalities~\ref{eq:sws-28}, the Christoffel tensor~\ref{eq:ani101-41} possesses a triplet of arbitrarily oriented, mutually orthogonal eigenvectors $\bm{U}^\mathrm{P}(\bm{n}^t) \perp \bm{U}^\mathrm{S1}(\bm{n}^t) \perp \bm{U}^\mathrm{S2}(\bm{n}^t)$. Therefore,  tensor $\bm{\Gamma}(\bm{n}^t)$ has to be scalar\cite{Alshitsetal1985},
\begin{equation} \label{eq:sws-29}
\bm{\Gamma}(\bm{n}^t) = V_t^2 \, \bm{I} \, ,
\end{equation}
where $\bm{I}$ is the $3 \times 3$ identity matrix. Equation~\ref{eq:sws-29} implies that certain equality-type constraints have to be imposed on the components of elastic stiffness tensor $\bm{c}$ for mere existence of a triple singularity. 

Perhaps the easiest way to determine these constraints is to orient the axis $\bm{x}_3$ of a local Cartesian coordinate frame along $\bm{n}^t$, making $\bm{n}^t = [0, ~ 0, ~ 1]$, and
\begin{equation} \label{eq:sws-27.6}
\bm{\Gamma}(\bm{n}^t) = \left( \begin{array}{c c c}
c_{55} & c_{45} & c_{35} \\
c_{45} & c_{44} & c_{34} \\
c_{35} & c_{34} & c_{33} \\
\end{array} \right) \! ,
\end{equation}
where the Voigt rule\cite{Musgrave1970, Tsvankin2001, Grechka2009EAGE} is applied to transition from the four-index notation of the components of fourth-rank stiffness tensor $\bm{c}$ to the two-index notation in matrix~\ref{eq:sws-27.6}.

The comparison of equations~\ref{eq:sws-29} and~\ref{eq:sws-27.6} reveals the sought equalities for the stiffness components
\begin{subnumcases}{\label{eq:sws-31}}
c_{34} = c_{35} = c_{45} = 0 \, , & \\
c_{33} = c_{44} = c_{55} = V_t^2 & 
\end{subnumcases}
in the selected local coordinate frame. Equations~\ref{eq:sws-31} clearly indicate the instability of triple singularities \cite{Alshitsetal1985} under an arbitrary triclinic perturbation of tensor $\bm{c}$, pointing to a low probability of finding triple  singularities in natural homogeneous solids like crystals. Setting these issues (revisited in the Discussion section) aside, triple singularities can be examined as purely theoretical or mathematical objects, just as anisotropic solids without conventional singularities \cite{AlshitsLothe1979} have been studied.

Such an investigation of triple singularities begins by recognizing ``a wholly arbitrary choice''\cite[][]{Musgrave1981} of unit polarization vector
\begin{equation}\label{eq:sws-33}
\bm{U}^t \equiv \bm{U}(\bm{n}^t, \, \varphi_1, \, \varphi_2) =
[ \sin \varphi_1 \cos \varphi_2, ~ \sin \varphi_1 \sin \varphi_2, ~ \cos \varphi_1] \, , \quad (0 \leq \varphi_1 \leq \pi, ~ 0 \leq \varphi_2 \leq 2 \, \pi) \, 
\end{equation}
as an eigenvector of tensor $\bm{\Gamma}(\bm{n}^t)$. Next, the general definition of group-velocity vector\cite{Fedorov1968, Musgrave1970}
\begin{equation} \label{eq:ani101-55}
\bm{g} = \bm{\Gamma}(\bm{U}) \bm{\cdot} \frac{\bm{n}}{V}
\end{equation}
and, specifically,
\begin{equation} \label{eq:sws-34}
\bm{g}^t \equiv \bm{g}(\bm{n}^t, \, \varphi_1, \, \varphi_2) = \bm{\Gamma}(\bm{U}^t)  
\bm{\cdot} \frac{\bm{n}^t}{V_t}
\end{equation}
yields the components of $\bm{g}^t$:
\begin{eqnarray}\label{eq:sws-35}
g_1^t \Bs = \frac{\sin \varphi_1}{V_t} \,
\Big\{ \cos \varphi_1 \, \big[ (c_{13} + c_{55}) \, \cos \varphi_2 + c_{36} \, \sin \varphi_2 \big]
+ \sin \varphi_1 \, \big[ c_{15} \, \cos^2 \varphi_2 + 
(c_{14} + c_{56}) \, \cos \varphi_2 \, \sin \varphi_2 
\nonumber \\
\Bs \hspace{30mm} + \, c_{46} \, \sin^2 \varphi_2 \, \big] \Big\} , 
\\
\label{eq:sws-36}
g_2^t \Bs = \frac{\sin \varphi_1}{V_t} \, 
\Big\{\cos \varphi_1 \, \big[ (c_{23} + c_{55}) \, \sin \varphi_2  + c_{36} \, \cos \varphi_2 \big]  
+ \sin \varphi_1 \, \big[ c_{56} \, \cos^2 \varphi_2 + (c_{25} + c_{46}) \, \cos \varphi_2 \, \sin \varphi_2 
\nonumber \\
\Bs \hspace{30mm} + \, c_{24} \, \sin^2 \varphi_2 \, \big] \Big\} ,
\\
\label{eq:sws-37}
g_3^t \Bs = V_t \, 
\end{eqnarray}
in our local coordinate frame. \vspace{0mm}

\begin{figure} 
	\centering
	\includegraphics[width=0.37\textwidth]{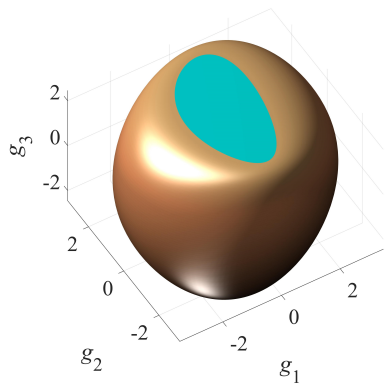}
	\caption{The P-wave group-velocity surface (copper) in triclinic model~\ref{eq:sws-38}. The planar portion of the surface displayed in cyan is the base $\bm{g}^Q$ of the internal refraction cone formed at triple singularity $\bm{n}^t = [0, \, 0, \, 1]$. 
	} 
	\label{fig:sws-07}
\end{figure}

Let us note the following. 
\begin{itemize} \vspace{-2mm}
	\item Vectors $\bm{g}(\bm{n}^t, \, \varphi_1, \, \varphi_2)$ fill a \emph{solid conical object} rather than trace the surface of an internal refraction cone, as analogous group-velocity vectors do at conventional (double) singularities. \vspace{-2mm}
	
	\item The component $g_3^t$ given by equation~\ref{eq:sws-37} is independent of the polarization angles $\varphi_1$ and $\varphi_2$; therefore, the base $\bm{g}^Q$ of the internal refraction cone is a \emph{plane} with the unit normal $\bm{n}^t$. \vspace{-2mm}
	
	\item The non-quadratic dependencies of the components $g_1^t$ and $g_2^t$ on sines and cosines of angles $\varphi_1$ and $\varphi_2$ in equations~\ref{eq:sws-35} and~\ref{eq:sws-36} generally result in a {quartic} rather than elliptic  (as for conventional singularities) shape of the base of an internal refraction cone, the feature of triple singularities illustrated in Figure~\ref{fig:sws-07} for a triclinic model that has the density-normalized stiffness tensor (in arbitrary units of velocity squared, as well as other stiffness tensors below) \vspace{-2mm} 
\end{itemize}
\begin{equation} \label{eq:sws-38}
\bm{c} = \left( \begin{array}{c c c c c c}
12 &  3 &  0 & 0.1 & 0.2 & 0.3 \\
~ &  10 &  5 & 0.4 & 0.5 & 0.6 \\
~ &  ~ &   6 & 0 & 0 & 0.4 \\
~ & \mathrm{SYM} & ~ &  6 & 0 & 0.7 \\
~ &  ~ &  ~ &           ~ & 6 & 0.8 \\
~ &  ~ &  ~ &           ~ & ~ & 4 \\
\end{array} \right) \! .
\end{equation}
Here ``SYM'' denotes symmetric part of the stiffness matrix.

It is straightforward to show that finite-size planar patches can be formed on group-velocity surfaces \emph{only} in conjunction with triple singularities. To see this, consider finite-size regular area of a group-velocity surface. Its flatness implies the zero Gaussian curvature, $\mathcal{K}_{\,g} = \mathcal{K}_{\,g}(\bm{n}) = 0$, for a range of the wavefront-normal vectors $\bm{n}$. Consequently, Gaussian curvature $\mathcal{K}_{\,p}$ of regular area of the corresponding slowness surface has to be infinite because\cite{GrechkaObolentseva1993, VavrycukYomogida1996}
\begin{equation} \label{eq:sws-22}
\mathcal{K}_{\,p} \, {\mathcal{K}_{\,g}} = \left(\frac{V}{|\bm{g}|} \right)^{\! 4}, 
\end{equation}
and the right side of equation~\ref{eq:sws-22} has to be finite to satisfy the elastic stability conditions. The equality $\mathcal{K}_{\,p} = \infty$ over a finite area of a slowness surface violates the differentiability of the Christoffel equation~\ref{eq:ani101-40} written in slownesses and implies that finite-size \emph{regular} region of a group-velocity surface cannot be planar, leaving singularities as the only remaining option. However, neither conventional conical nor intersection singularities yield flat areas of the group-velocity surfaces; the former, occurring along isolated wavefront-normal directions, map themselves onto elliptical lines of the zero Gaussian curvature\cite{ShuvalovEvery1997} $\mathcal{K}_{\,g} = 0$ rather than onto finite-size areas; the latter, shaped as circles in transversely isotropic media, do give rise to the zero Gaussian curvature areas but these areas, either conical or cylindric, are not planar (see an example in section~\ref{ch:sws-triple-VTI}).

\section{The minimax property} \label{ch:sws-triple-minimax} 

The phase velocity $V_t$ defined by equations~\ref{eq:sws-28} possesses a remarkable property\cite{AlshitsLothe1979a, Alshits2003}
\begin{equation} \label{eq:sws-ext01}
V_t = \min \limits_{(\bm{n})} V_\mathrm{P}(\bm{n}) = \max \limits_{(\bm{n})} V_\mathrm{S2}(\bm{n}) \, .
\end{equation} 
In a word, velocity $V_t$ in the direction of triple singularity $\bm{n}^t$ is equal to both \vspace{-2mm}
	\begin{itemize} 
		\item the global minimum of the outermost phase-velocity sheet $V_\mathrm{P}(\bm{n})$ and \vspace{-2mm}
		\item the global maximum of the innermost phase-velocity sheet $V_\mathrm{S2}(\bm{n})$. \vspace{-1.5mm}
	\end{itemize}

This statement is a direct consequence of inequality\cite{AlshitsLothe1979}
\begin{equation} \label{eq:sws-ext02}
V_\mathrm{P} \! \left( \bm{n}^{a} \right) \geq V_\mathrm{S2} \! \left(\bm{n}^{b} \right) ~ \forall ~ \, \bm{n}^{a} ~ \text{and} ~ \bm{n}^{b} \, .
\end{equation}
Indeed, in accordance with equations~\ref{eq:sws-28}, the wavefront normal direction $\bm{n}^{a} = \bm{n}^{b} = \bm{n}^t$ is exactly where 
\begin{equation} \label{eq:sws-ext02a}
V_\mathrm{P}(\bm{n}^t) = V_\mathrm{S2}(\bm{n}^t) = V_t \, ,
\end{equation}
turning inequality~\ref{eq:sws-ext02} into equality. 

The minimax property~\ref{eq:sws-ext01} leads to another noteworthy corollary, \vspace{-1.5mm}
\begin{quote}
	If an anisotropic solid possesses $N$ distinct triple singularities $\bm{n}^{t_1}, \, \ldots, \, \bm{n}^{t_N}$, the phase velocities along \emph{all of them} are equal,
\end{quote}	 \vspace{-1mm}	
\begin{equation} \label{eq:sws-ext03}
V \! \left( \bm{n}^{t_1} \right) = \ldots = V \! \left( \bm{n}^{t_N} \right) = V_t \, .
\end{equation} 

\section{The maximum number of triple singularities} \label{ch:sws-triple-maxnum} 

To establish the maximum number $N$ of triple singularities in equation~\ref{eq:sws-ext03}, we rewrite equation~\ref{eq:sws-29} as
\begin{equation} \label{eq:sws-ext04}
\bm{\Gamma}(\bm{p}^{t}) = \bm{I} ,
\end{equation}
where 
\begin{equation} \label{eq:sws-ext05}
\bm{p}^{t} = \frac{\bm{n}^t}{V_t} \, 
\end{equation}
is the slowness vector in at a triple singularity. Next, we recognize equation~\ref{eq:sws-ext04} as a system of six quadratic equations for three unknown components of vector $\bm{p}^{t}$. 

In general, system~\ref{eq:sws-ext04} is incompatible because it contains six equations for only three unknowns. Its incompatibility is consistent though with equality-type constraints~\ref{eq:sws-31} on the stiffness coefficients that have to be imposed in the special coordinate frame for the triple singularity directed at $\bm{n}^t = [0, \, 0, \, 1]$. Bearing the necessity of these constraints in mind, we split system~\ref{eq:sws-ext04} into two subsystems, each including three equations for three elements ${\Gamma}_{i j}(\bm{p}^{t})$ of the Christoffel matrix $\bm{\Gamma}(\bm{p}^{t})$. The first subsystem of three polynomial equations
\begin{equation} \label{eq:sws-ext06}
{\Gamma}_{i j}(\bm{p}^{t}) = 
\def\arraystretch{1.2}
\left\{ \begin{array}{rl} 
0 & \! \text{if }\, i \neq j \, , \\
1 & \! \text{if }\, i = j \, , \\
\end{array}
\right.
\def\arraystretch{1.0}
\end{equation}
in which $i = j$ for at least one of the equations, could be solved for the three components of $\bm{p}^{t}$; whereas the remaining subsystem, 
\begin{equation} \label{eq:sws-ext07}
{\Gamma}_{i' j'}(\bm{p}^{t}) = 
\def\arraystretch{1.2}
\left\{ \begin{array}{rl} 
0 & \! \text{if }\, i' \neq j' \, , \\
1 & \! \text{if }\, i' = j' \, , \\
\end{array}
\right.
\def\arraystretch{1.0}
\end{equation}
comprising three equations for three elements ${\Gamma}_{i' j'}(\bm{p}^{t})$ of matrix $\bm{\Gamma}(\bm{p}^{t})$ with pairs of indexes $(i', \, j') \neq (i, \, j)$, would provide constraints on stiffnesses $\bm{c}$ analogous to those given by equations~\ref{eq:sws-31}.

If three equations~\ref{eq:sws-ext06} are compatible and non-degenerative, the maximum number of their real-valued roots is given by  B\'{e}zout's theorem\cite{Weisstein2003} as a product of the degrees of equations, that is, \mbox{$2 \times 2 \times 2 = 8$.} We note that real-valued roots of equations~\ref{eq:sws-ext06} always appear in centrally symmetric pairs $\bm{p}^{t}$ and $-\bm{p}^{t}$ because the left sides ${\Gamma}_{i j}(\bm{p}^{t})$ of equations~\ref{eq:sws-ext06} are homogeneous functions of degree 2 of the components of vector $\bm{p}^{t}$; hence, the maximum number $N$ of distinct, non-centrally symmetric triple singularities is equal to 
\begin{equation} \label{eq:sws-ext08}
N = 4 \, .
\end{equation} 

In general, there is no guarantee that triplets of equations~\ref{eq:sws-ext07}, which have to be satisfied for all real-valued roots $\bm{p}^{t_k}$ of subsystem~\ref{eq:sws-ext06}, yield a compatible set of constraints on the stiffness coefficients $\bm{c}$ similar to equations~\ref{eq:sws-31}. If subsystem~\ref{eq:sws-ext07} happens to be self-contradictory for some roots $\bm{p}^{t_k}$, the directions of those vectors $\bm{p}^{t_k}$ would not obey the criteria for valid triple singularities. In the next section, however, we demonstrate that the theoretically derived maximum~\ref{eq:sws-ext08} is realizable~| we are going to construct a set of orthorhombic solids possessing exactly $N = 4$ triple singularities.

\section{Orthotropy} 

Even though equalities~\ref{eq:sws-31}a, required for the presence of a triple singularity at the vertical, are automatically satisfied in orthorhombic media\cite{Fedorov1968, Musgrave1970}, constraints~\ref{eq:sws-31}b still have to be imposed, implying that only orthorhombic solids of a special kind can have triple singularities. The orthorhombic symmetry simplifies equations~\ref{eq:sws-34}~--~\ref{eq:sws-37} to
\begin{equation}\label{eq:sws-39}
\def\arraystretch{2.2}
\bm{g}^t = \bm{g}(\bm{n}^t, \, \varphi_1, \, \varphi_2) = \left[ \begin{array}{c}
\dfrac{c_{13} + c_{55}}{2 \, V_t} \, \sin 2 \, \varphi_1 \, \cos \varphi_2  \\
\dfrac{c_{23} + c_{55}}{2 \, V_t} \, \sin 2 \, \varphi_1 \, \sin \varphi_2  \\
V_t
\end{array} \right] \! ,
\def\arraystretch{1.0}
\end{equation}
leading to an elliptic rather than quartic internal refraction cone with the semi-axes 
\begin{equation}\label{eq:sws-40}
g_1^t \! \left( \bm{n}^t, \, \frac{\pi}{4}, \, 0 \right) = 
\frac{1}{2} \left(\frac{c_{13}}{V_t} + V_t \right)
\end{equation}
and
\begin{equation}\label{eq:sws-41}
g_2^t \! \left( \bm{n}^t, \, \frac{\pi}{4}, \, \frac{\pi}{2} \right) = 
\frac{1}{2} \left(\frac{c_{23}}{V_t} + V_t \right) \! .
\end{equation}

\begin{figure}[t]
	\centering
	\begin{tabular}{c c}
		~ \, \qquad (a) & \qquad (b) \\
		\includegraphics[height=0.30\textwidth]{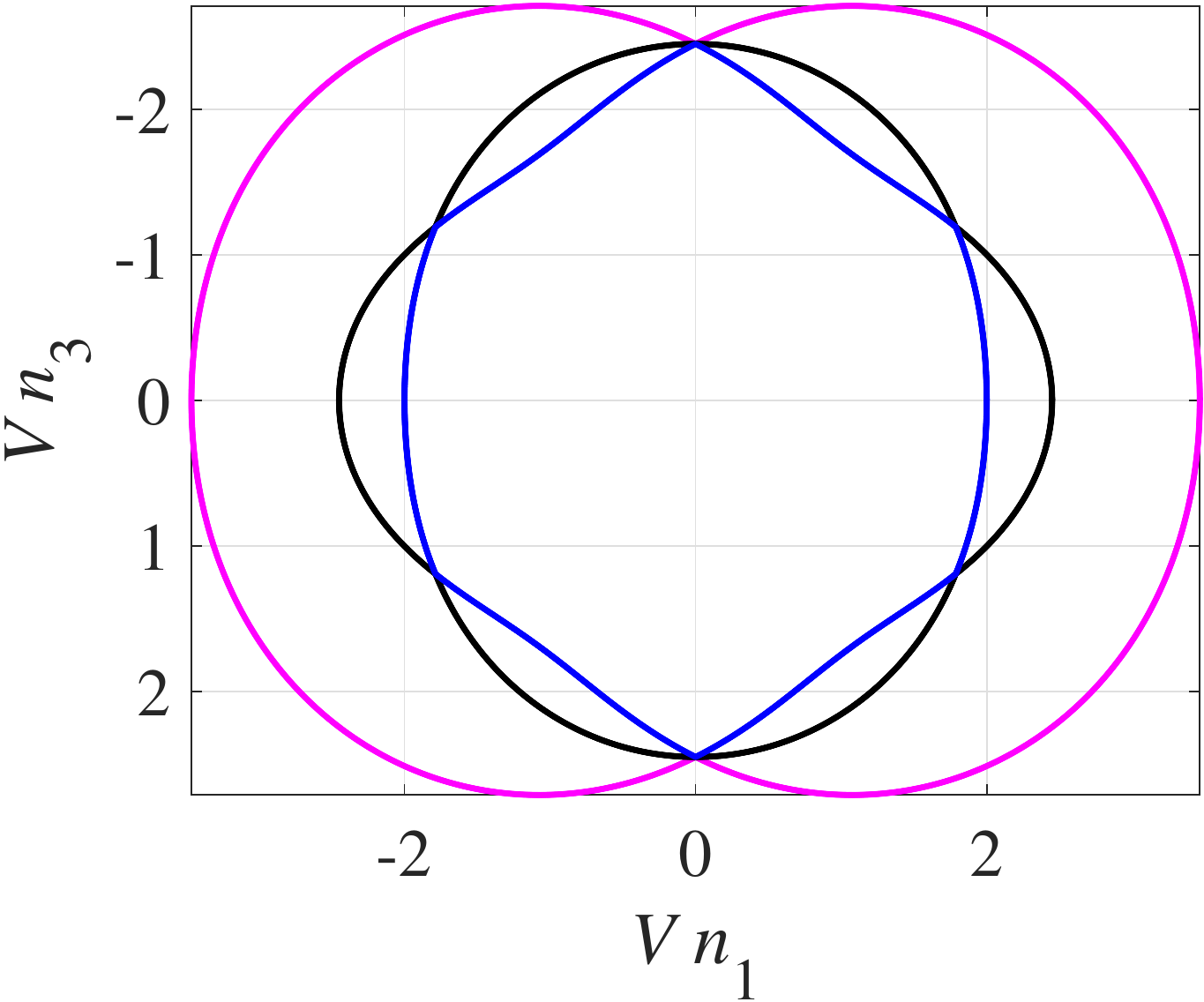} &
		\includegraphics[height=0.31\textwidth]{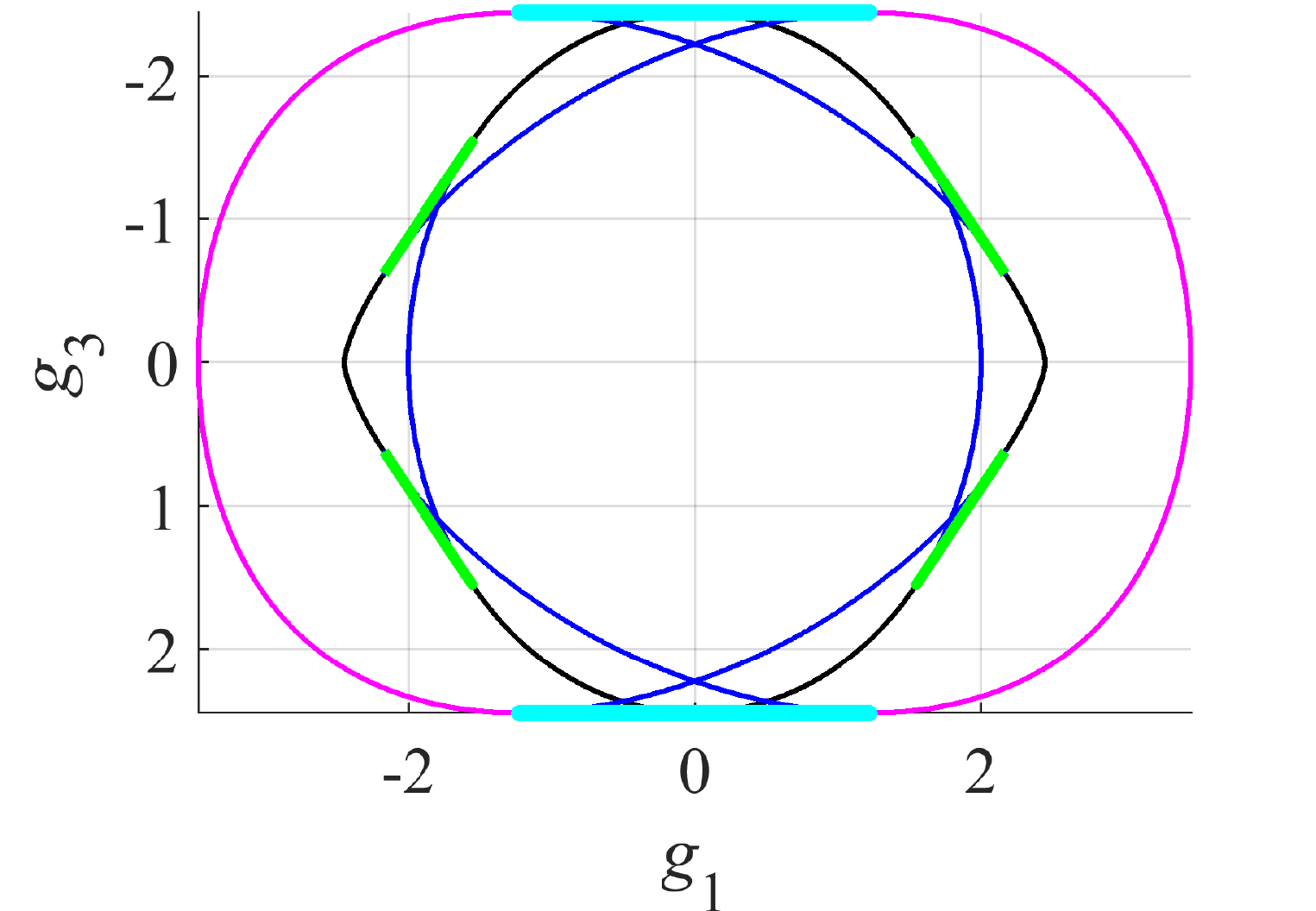} \\
		~ \, \qquad (c) & \qquad (d) \\
		\includegraphics[height=0.30\textwidth]{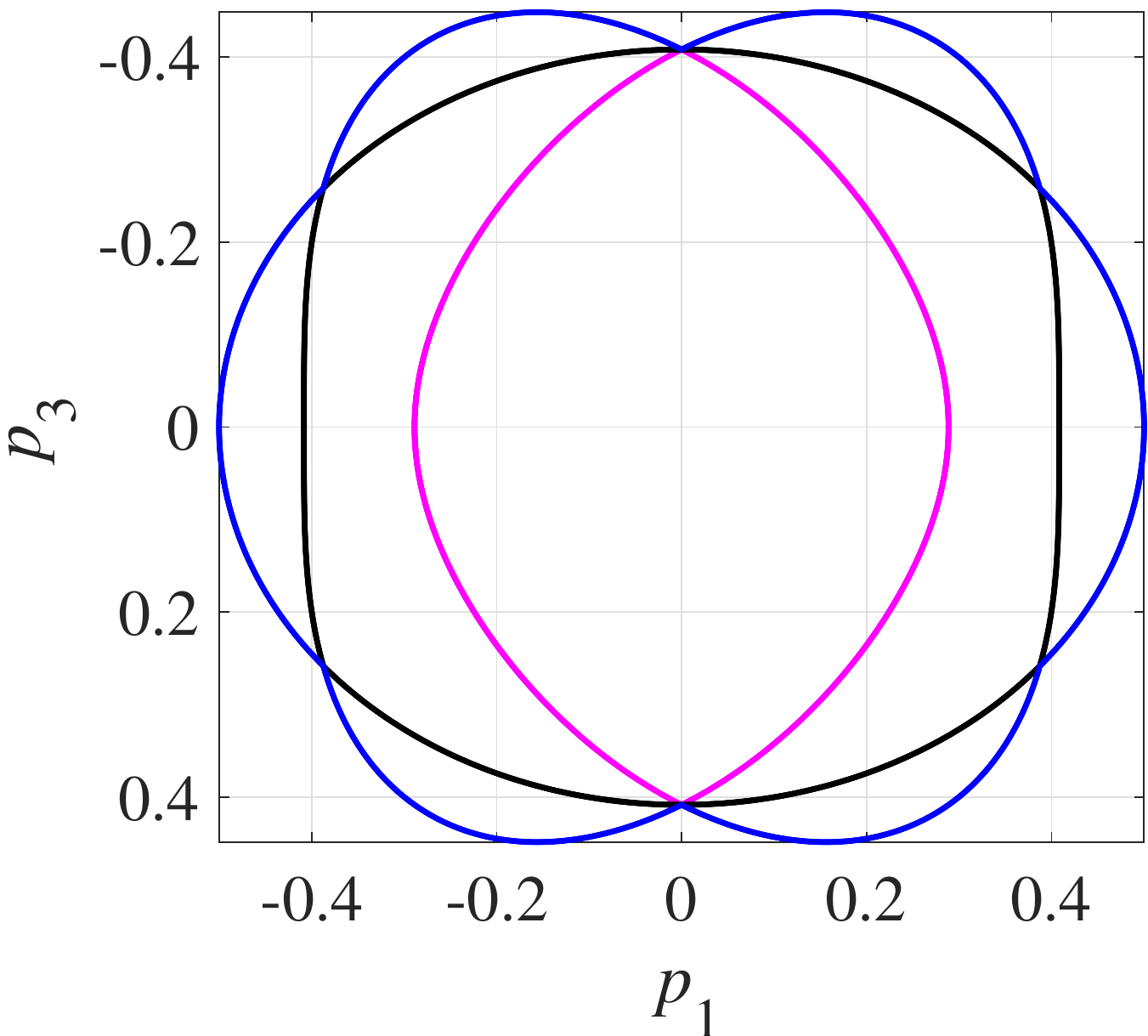} &
		\includegraphics[height=0.30\textwidth]{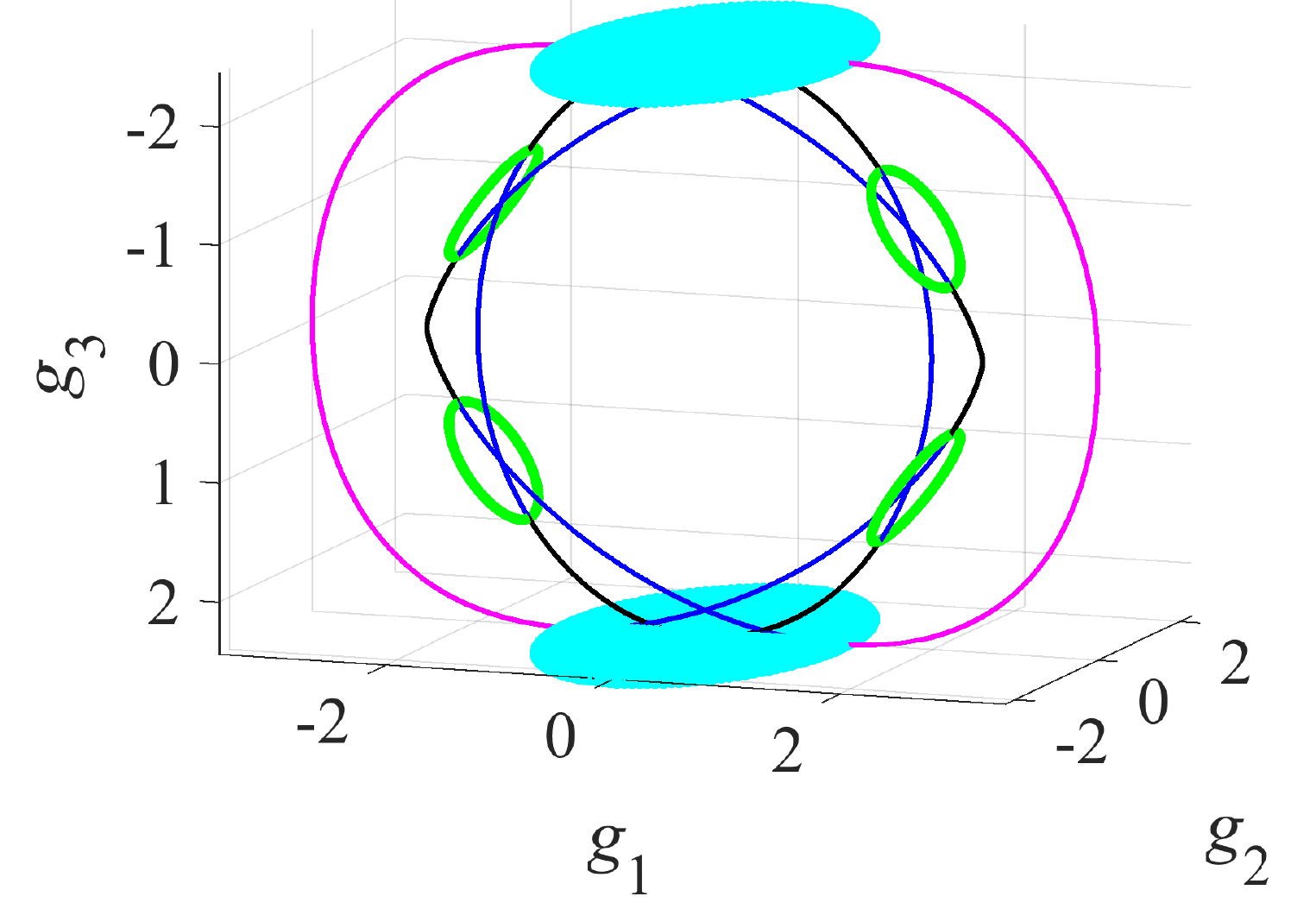} \\
	\end{tabular}
	\caption{(a) Phase-velocity, (b) group-velocity, and (c) slowness surfaces of the P- (magenta), S$_1$- (black), and S$_2$-waves (blue) in the $[\bm{x}_1, \, \bm{x}_3]$ plane of orthorhombic model~\ref{eq:sws-42}. Figure~(d) presents a 3D view of (b), displaying four bases $\bm{g}^E$ (the green lines) of the internal refraction cones related to conical singularities $\bm{n}^s = [\pm 0.832, \, 0, \, \pm 0.555]$ in the symmetry plane $[\bm{x}_1, \, \bm{x}_3]$ and two bases $\bm{g}^Q$ (the filled cyan ellipses) related to triple singularities at $\bm{n}^t = [0, \, 0, \, \pm 1]$.
	}
	\label{fig:sws-08} \vspace{-3mm}
\end{figure}

Figure~\ref{fig:sws-08}, computed for an orthorhombic medium characterized by the stiffness matrix
\begin{equation} \label{eq:sws-42}
\bm{c} = \left( \begin{array}{c c c c c c}
12 &  3 &  0 & 0 & 0 & 0 \\
~ &  10 &  5 & 0 & 0 & 0 \\
~ &  ~ &   6 & 0 & 0 & 0 \\
~ & \mathrm{SYM} & ~ &  6 & 0 & 0 \\
~ &  ~ &  ~ &           ~ & 6 & 0 \\
~ &  ~ &  ~ &           ~ & ~ & 4 \\
\end{array} \right) \! 
\end{equation}
obeying constraints~\ref{eq:sws-31}b, displays the bases of internal refraction cones corresponding to both triple (cyan) and shear-wave conical (green) singularities that can coexist in the same model.

\begin{figure}
	\centering
	\begin{tabular}{c c}
		~ \, \, \qquad (a) & ~ \, \qquad (b) \\
		\includegraphics[height=0.30\textwidth]{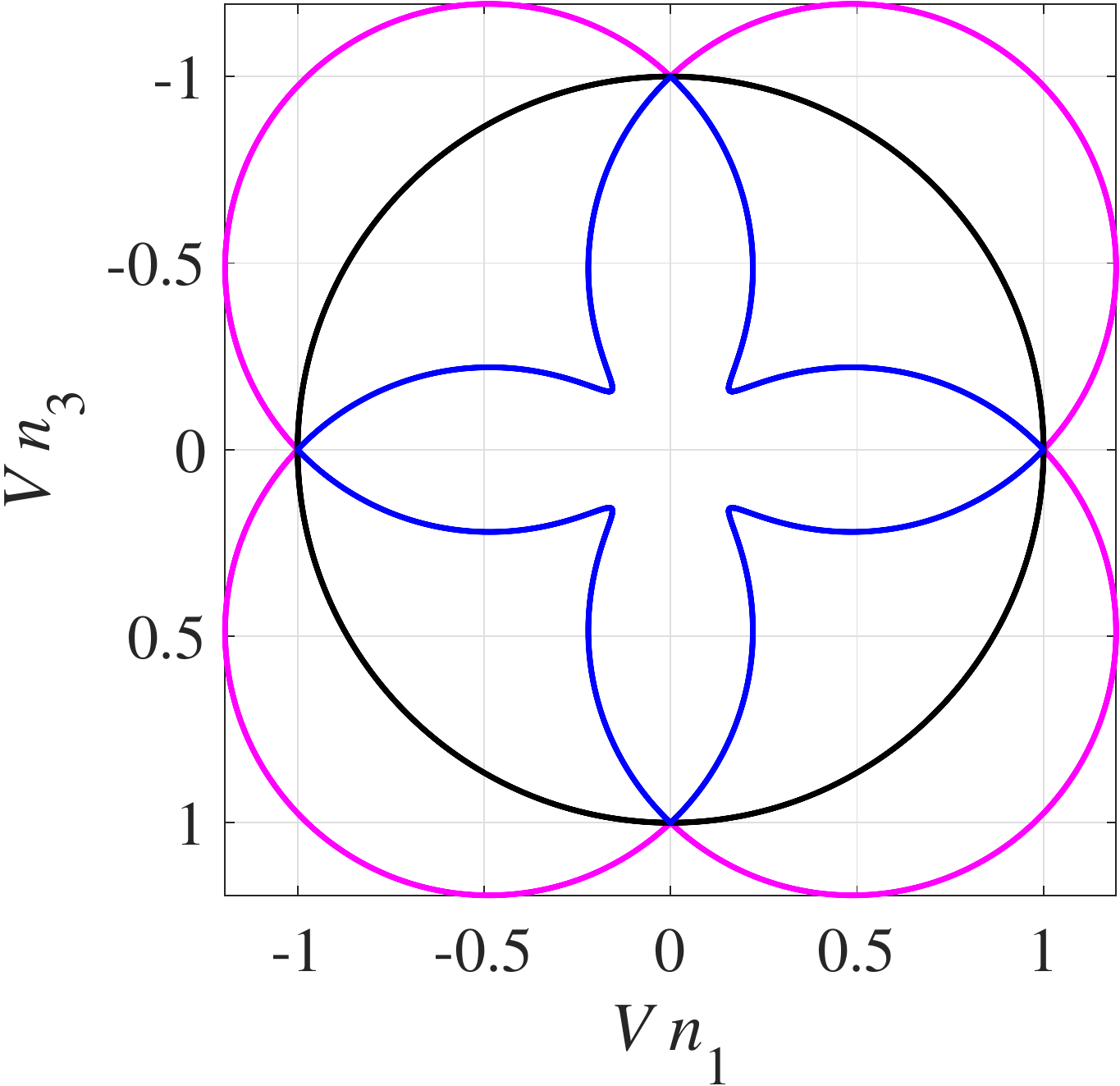} &
		\includegraphics[height=0.30\textwidth]{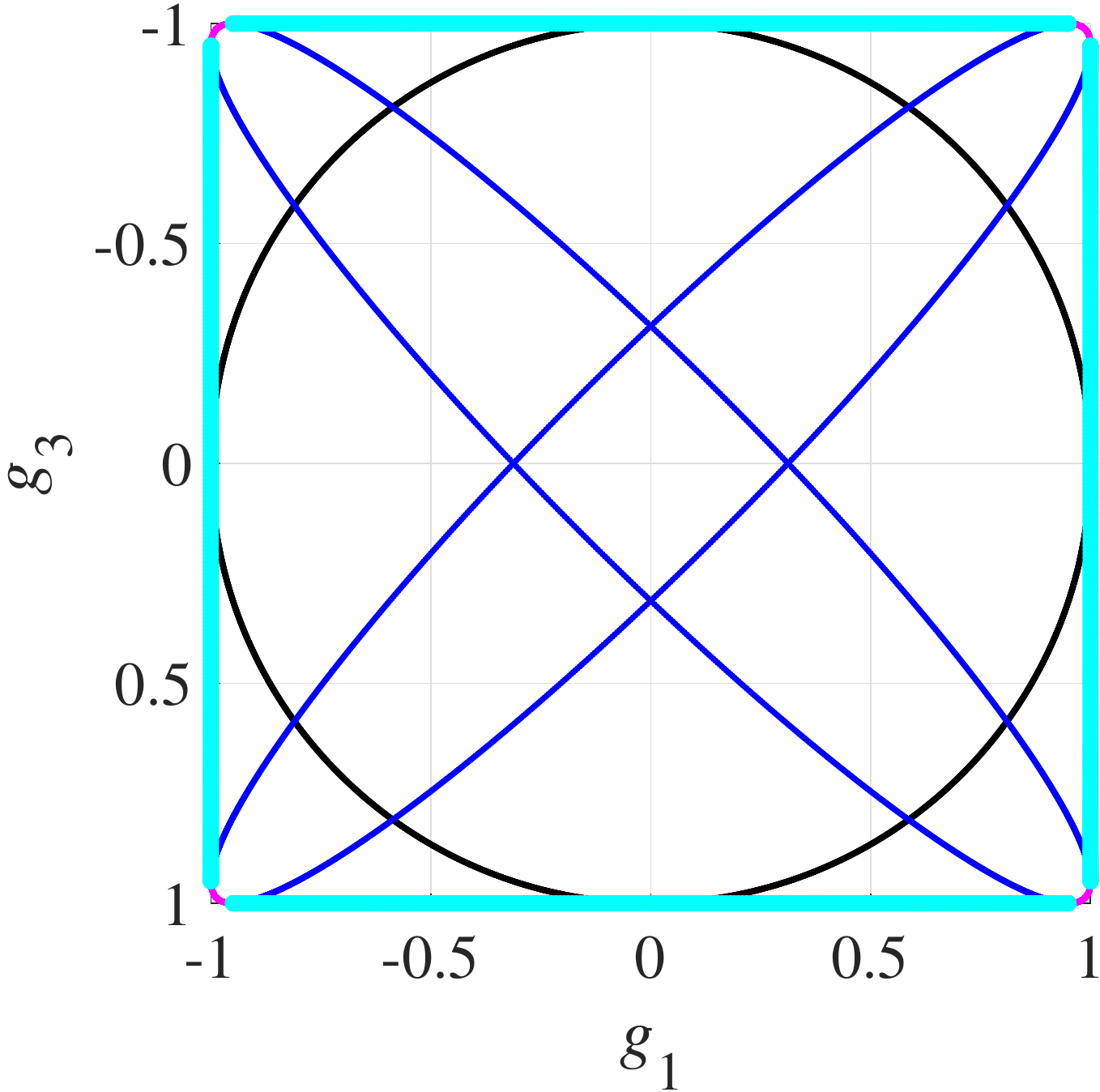} \\
		~ \, \qquad (c) & ~ \, \qquad (d) \\
		\includegraphics[height=0.30\textwidth]{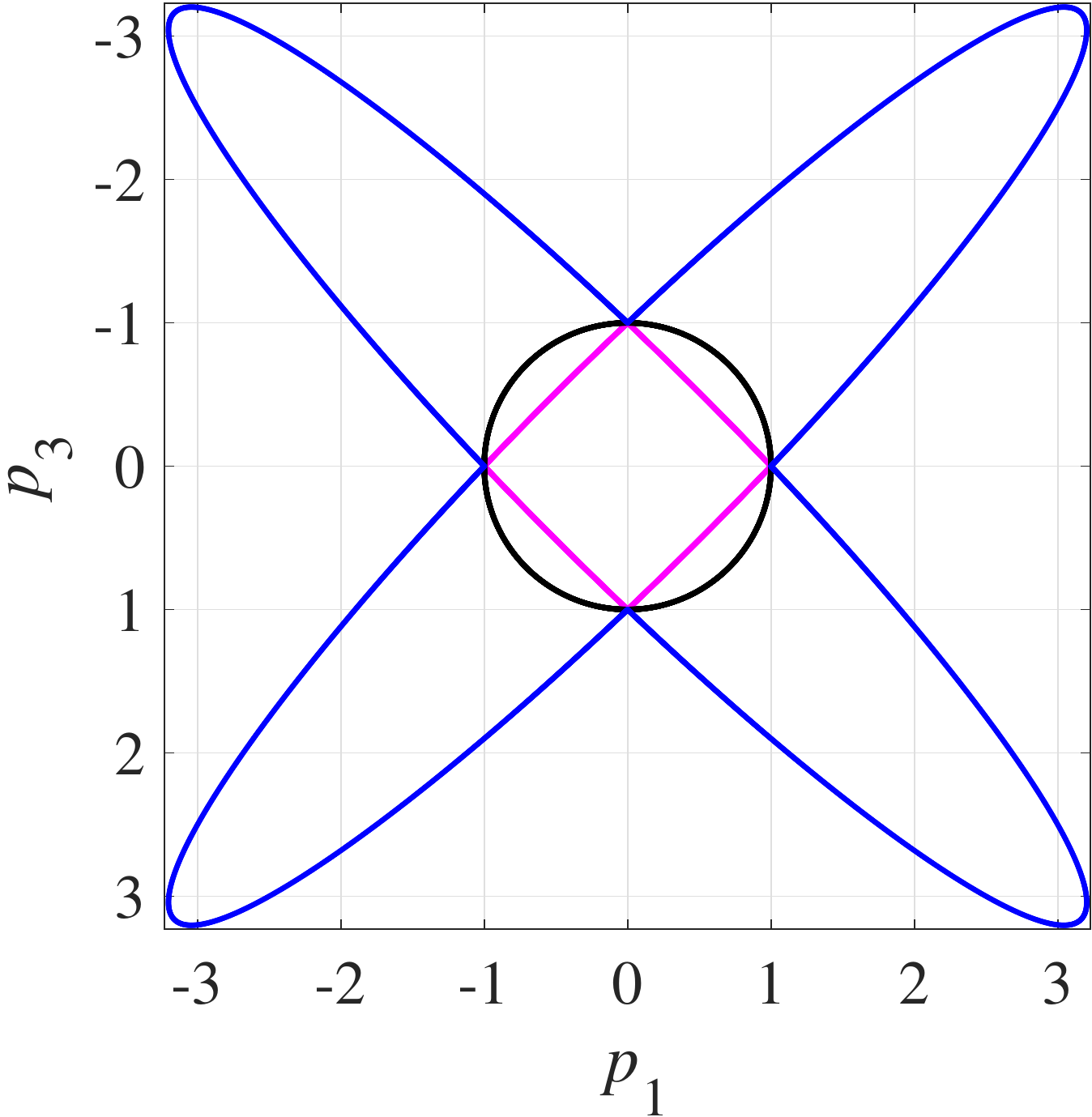} &
		\includegraphics[height=0.31\textwidth]{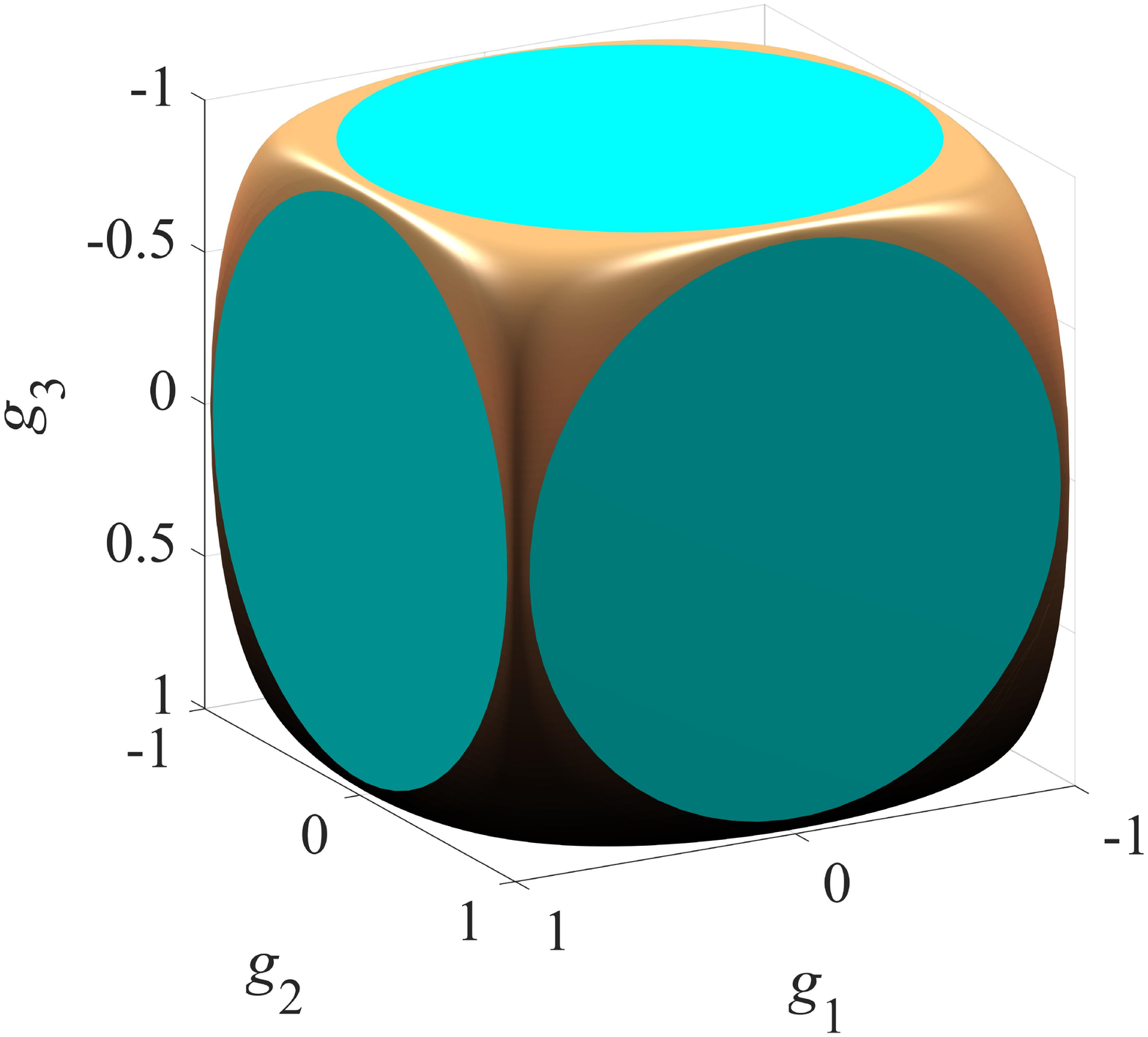} \\
	\end{tabular}
	\caption{(a)~--~(c) Same as Figures~\ref{fig:sws-08}a~--~\ref{fig:sws-08}c but for cubic stiffness matrix~\ref{eq:sws-42.1}. Model~\ref{eq:sws-42.1} has no conical singularities in the $[\bm{x}_1, \, \bm{x}_3]$ symmetry plane; the bases $\bm{g}^E$ of internal refraction cones at out of plane singularities $\bm{n}^s = [\pm 1, \, \pm 1, \, \pm 1]/\sqrt{3}$ are not shown. As in Figure~\ref{fig:sws-08}, the filled cyan circles in (d) and their projections onto the plane $[\bm{x}_1, \, \bm{x}_3]$ in (b) are the bases $\bm{g}^Q$ of internal refraction cones corresponding to triple singularities. 3D surface in (d) is the outer (P-wave) sheet of the group-velocity surface.
	}
	\label{fig:sws-08.1} \vspace{0mm} 
\end{figure}

A more intellectually pleasing and certainly more instructive example, illustrating the findings of section~\ref{ch:sws-triple-minimax}, is shown in Figure~\ref{fig:sws-08.1} for a cubic (a special case of orthorhombic) model
\begin{equation} \label{eq:sws-42.1}
\bm{c} = \left( \begin{array}{c c c c c c}
1 &  0.9 &  0.9 & 0 & 0 & 0 \\
~ &  1 & 0.9 & 0 & 0 & 0 \\
~ &  ~ &  1 & 0 & 0 & 0 \\
~ & \mathrm{SYM} & ~ & 1 & 0 & 0 \\
~ &  ~ &  ~ &           ~ & 1 & 0 \\
~ &  ~ &  ~ &           ~ & ~ & 1 \\
\end{array} \right) \! .
\end{equation}
As can be easily verified, model~\ref{eq:sws-42.1} possesses triple singularities along all three coordinate axes $\bm{n}^{t_1} = [\pm 1, \, 0, \, 0]$, $\bm{n}^{t_2} = [0, \, \pm 1, \, 0]$, and $\bm{n}^{t_3} = [0, \, 0, \, \pm 1]$. According to equations~\ref{eq:sws-ext03}, the phase velocities have to coincide for all singular directions $\bm{n}^{t_i}$ $(i = 1, \, 2, \, 3)$, 
\begin{equation} \label{eq:sws-42.3}
V \! \left( \bm{n}^{t_1} \right) = V \! \left( \bm{n}^{t_2} \right) = V \! \left( \bm{n}^{t_3} \right) ,
\end{equation}
and Figure~\ref{fig:sws-08.1}a displays equality of the two velocities, $V \! \left( \bm{n}^{t_1} \right) = V \! \left( \bm{n}^{t_3} \right) = 1$. 

\begin{figure}
	\centering
	\includegraphics[width=0.4\textwidth]{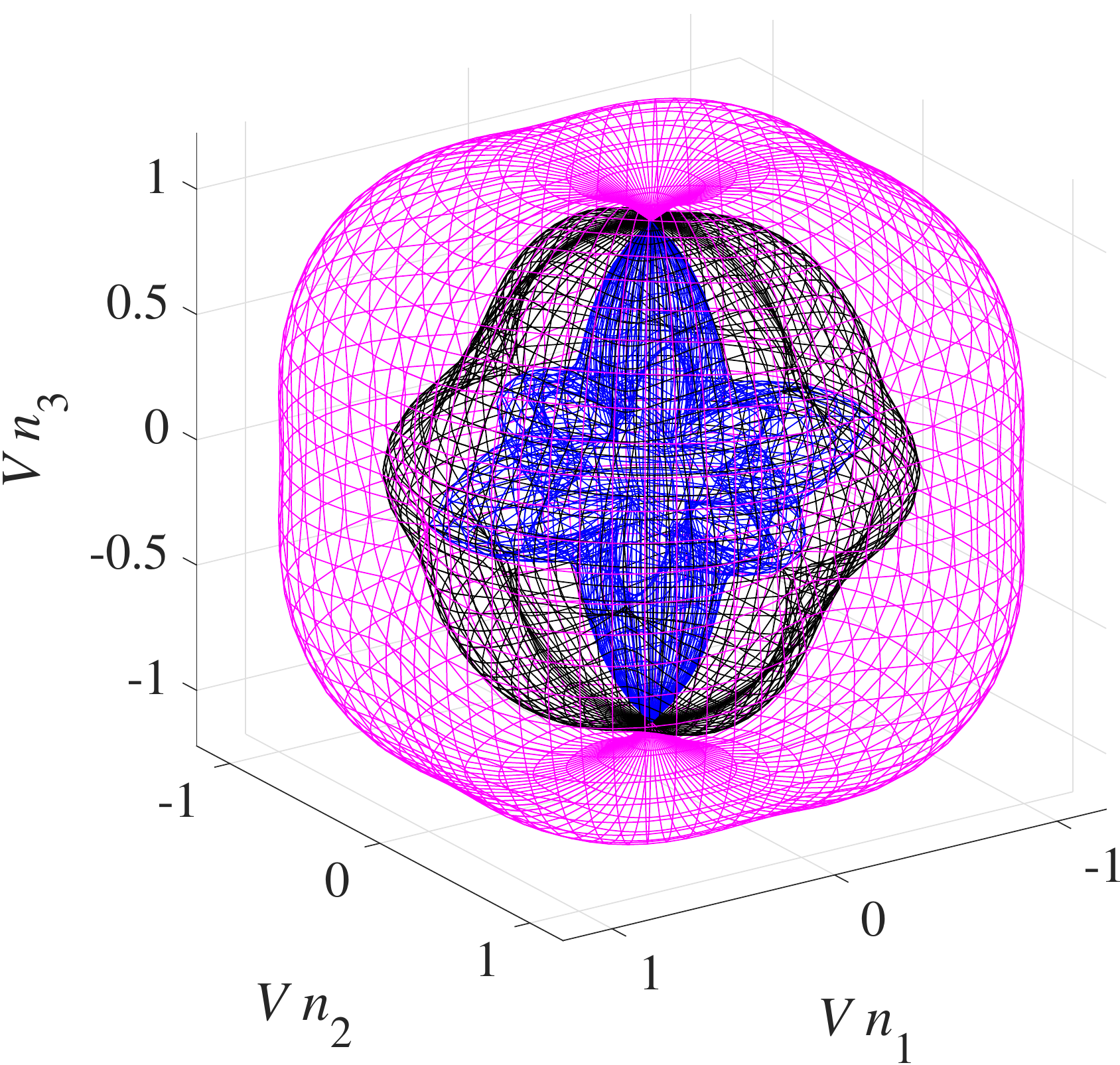}
	\caption{Phase-velocity sheets of the P- (magenta), S$_1$- (black), and S$_2$-waves (blue). The phase-velocity sheet of the S$_1$-wave is not a sphere even though its cross-section in the $[\bm{x}_1, \, \bm{x}_3]$ plane \emph{is} a circle (Figure~\ref{fig:sws-08.1}a).
	}
	\label{fig:sws-08.2}
\end{figure}

Because $c_{44} = c_{66} = 1$ in stiffness matrix~\ref{eq:sws-42.1}, the S$_1$- or SH-wave exhibits kinematically isotropic behavior in the vertical symmetry plane $[\bm{x}_1, \, \bm{x}_3]$ (the black circles in Figure~\ref{fig:sws-08.1}a~--~\ref{fig:sws-08.1}c), providing a convenient baseline for the visual confirmation of inequality~\ref{eq:sws-ext02}. Figure~\ref{fig:sws-08.1}a shows that the phase velocity of the P-wave (magenta) is greater than the phase velocity of the S$_2$-wave (blue) everywhere except for directions $\bm{n}^{t_1}$ and $\bm{n}^{t_3}$, along which the two phase velocities coincide, 
\begin{equation} \label{eq:sws-42.4}
V_\mathrm{P}(\bm{n}^{t_1}) = V_\mathrm{S2}(\bm{n}^{t_1}) = V_\mathrm{P}(\bm{n}^{t_3}) = V_\mathrm{S2}(\bm{n}^{t_3}) = 1 \, .
\end{equation}
The same relationship holds for the entire P- and S$_2$-wave phase-velocity sheets in 3D (Figure~\ref{fig:sws-08.2}), not just for their cross-sections by the $[\bm{x}_1, \, \bm{x}_3]$ plane. 

The equality of off-diagonal stiffness elements $c_{12} = c_{13} = c_{23}$ in matrix~\ref{eq:sws-42.1} degenerates quartic bases $\bm{g}^Q$ of all internal refraction cones to circles (see equations~\ref{eq:sws-40} and~\ref{eq:sws-41}). These bases form \emph{planar patches} at the group-velocity surfaces, shown in cyan in Figure~\ref{fig:sws-08.1}d for the P-wave group-velocity sheet, exhibiting shape of a delightful dice. 

Model~\ref{eq:sws-42.1} has three triple singularities. Next, we show that the maximum number of triple singularities~| four~|  is also attainable. 

The system of six equations~\ref{eq:sws-ext04} in orthorhombic media reads 
\begin{subnumcases}{\label{eq:sws-42.5}}
\Gamma_{11}(\bm{p}^{t}) = c_{11} \, (p^t_1)^2 + c_{66} \, (p^t_2)^2 + c_{55} \, (p^t_3)^2 = 1 \, , & \\
\Gamma_{12}(\bm{p}^{t}) = (c_{12} + c_{66}) \, p^t_1 \, p^t_2 = 0 \, , & \\			
\Gamma_{13}(\bm{p}^{t}) = (c_{13} + c_{55}) \, p^t_1 \, p^t_3 = 0 \, , & \\		
\Gamma_{22}(\bm{p}^{t}) = c_{66} \, (p^t_1)^2 + c_{22} \, (p^t_2)^2 + c_{44} \, (p^t_3)^2 = 1 \, , & \\
\Gamma_{23}(\bm{p}^{t}) = (c_{23} + c_{44}) \, p^t_2 \, p^t_3 = 0 \, , & \\		
\Gamma_{33}(\bm{p}^{t}) = c_{55} \, (p^t_1)^2 + c_{44} \, (p^t_2)^2 + c_{33} \, (p^t_3)^2 = 1 \, . & 
\end{subnumcases}
We satisfy equations~\ref{eq:sws-42.5}b, \ref{eq:sws-42.5}c, and \ref{eq:sws-42.5}e identically by imposing constraints on the stiffness coefficients
\begin{subnumcases}{\label{eq:sws-42.6}}
c_{12} = -c_{66} \, , & \\			
c_{13} = -c_{55} \, , & \\		
c_{23} = -c_{44} \, & 
\end{subnumcases}
analogous to equations~\ref{eq:sws-31}a. Then the three remaining equations~\ref{eq:sws-42.5}a, \ref{eq:sws-42.5}d, and \ref{eq:sws-42.5}f, linear in the squared slowness components $(p^t_i)^2$ $(i = 1, \, 2, \, 3)$, can be solved analytically, yielding eight roots $\bm{p}^t$ with the components
\begin{subnumcases}{\label{eq:sws-42.7}}
p^t_1 = \pm \frac{1}{d} \, \sqrt{c_{22} \, c_{33} - c_{22} \, c_{55} - c_{33} \, c_{66} + c_{44} \, (c_{55} + c_{66} - c_{44})} \, , & \\			
p^t_2 = \pm \frac{1}{d} \, \sqrt{c_{11} \, c_{33} - c_{11} \, c_{44} - c_{33} \, c_{66} + c_{55} \, (c_{44} + c_{66} - c_{55})} \, , & \\			
p^t_2 = \pm \frac{1}{d} \, \sqrt{c_{11} \, c_{22} - c_{11} \, c_{44} - c_{22} \, c_{55} + c_{66} \, (c_{44} + c_{55} - c_{66})} \, , & 
\end{subnumcases}
where the denominators
\begin{equation} \label{eq:sws-43.7}
d = \sqrt{c_{11} \, c_{22} \, c_{33} + 2 \, c_{44} \, c_{55} \, c_{66} - c_{11} \, c_{44}^2 - c_{22} \, c_{55}^2 - c_{33} \, c_{66}^2} \, .
\end{equation}
As can be directly verified, the phase velocities $V_t$ coincide for all roots~\ref{eq:sws-42.7},
\begin{eqnarray} \label{eq:sws-43.8}
V_t \equiv \frac{1}{|\bm{p}^t|} = d \! \Bs \Big[ \, 2 \, (c_{44} \, c_{55} + c_{44} \, c_{66} + c_{55} \, c_{66} - c_{11} \, c_{44} - c_{22} \, c_{55} - c_{33} \, c_{66})
\nonumber \\
\Bs \, \, \, + \, c_{11} \, c_{22} + c_{11} \, c_{33} + c_{22} \, c_{33} - c_{44}^2 - c_{55}^2 - c_{66}^2 \Big]^{-1/2} \, ,
\end{eqnarray}
confirming our theoretical assertion~\ref{eq:sws-ext03}.

\begin{figure}[h]
	\centering
	\begin{tabular}{c c}
		~ \quad (a) & ~ \quad (b) \\
		\includegraphics[height=0.4\textwidth]{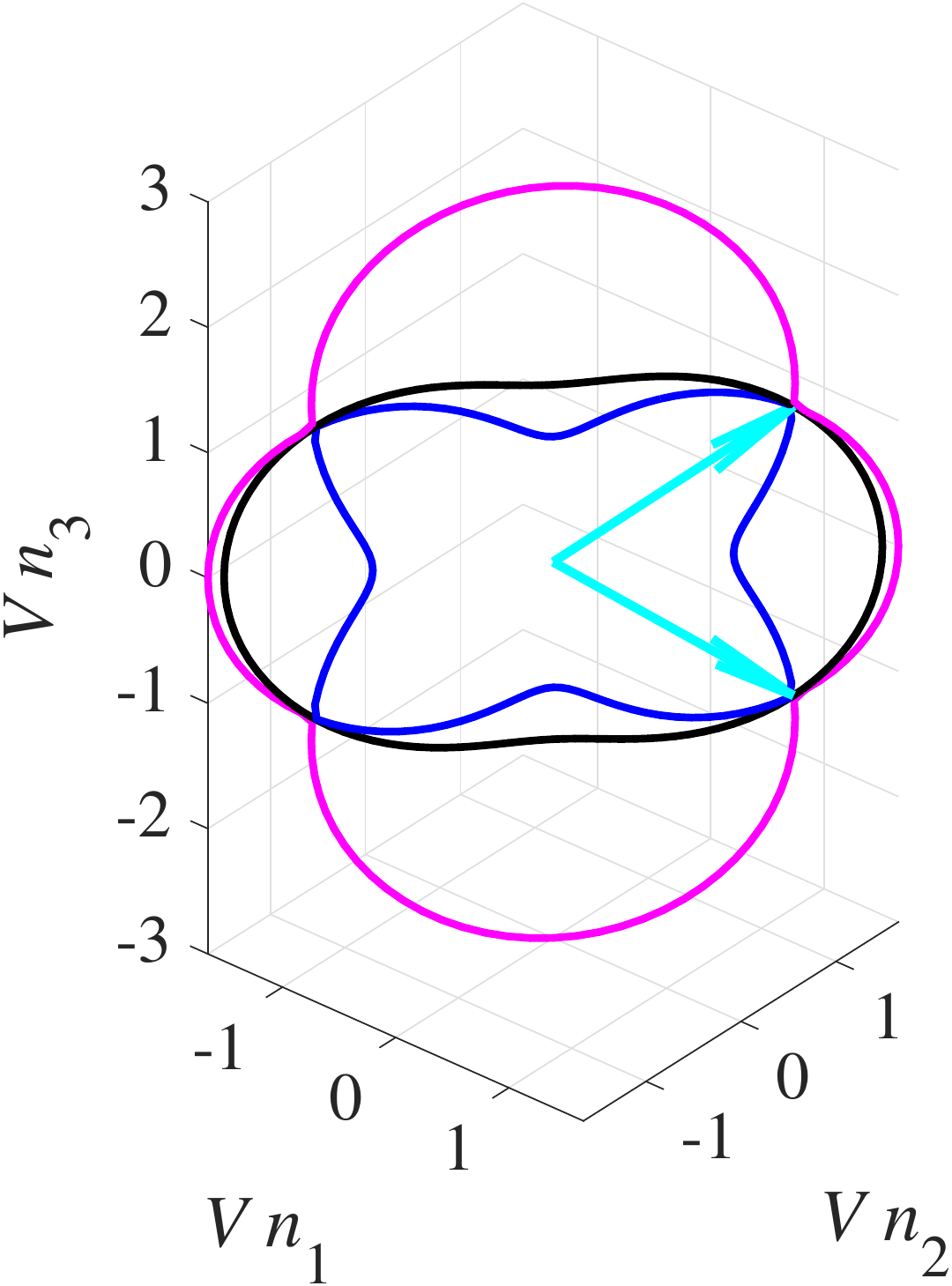} &
		\includegraphics[height=0.43\textwidth]{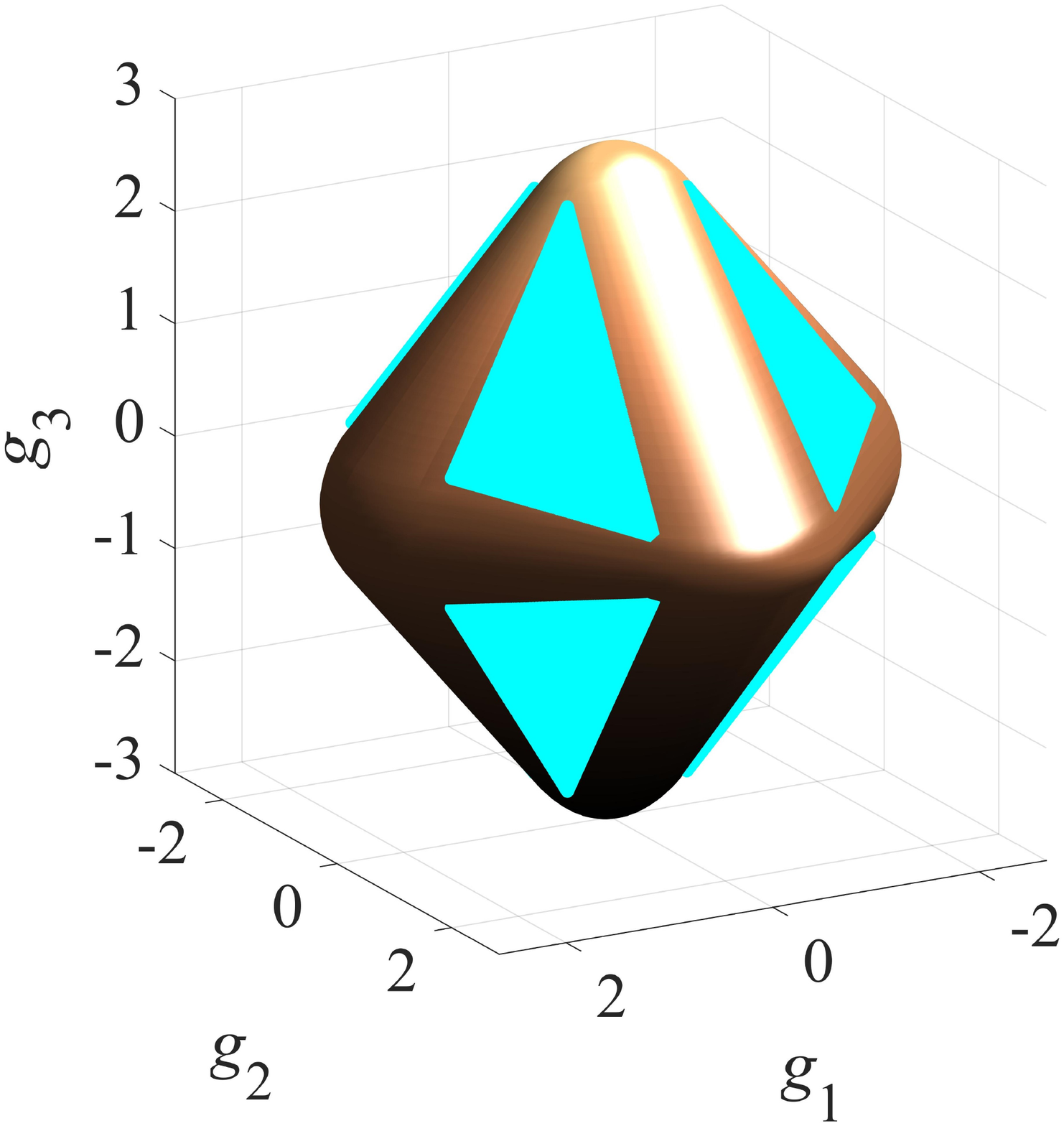} \\
	\end{tabular} \vspace{-2mm}
	\caption{(a) Cross-sections of the phase-velocity surfaces of the P- (magenta), S$_1$- (black), and S$_2$-waves (blue) by the vertical plane oriented at $45^\circ$ with respect to the horizontal coordinate axes $\bm{x}_1$ and $\bm{x}_2$ and (b) the P-wave sheet of the group-velocity surface for orthorhombic model~\ref{eq:sws-43.9}. The cyan arrows in (a) point to triple singularities $\bm{n}^{t_1}$ and $\bm{n}^{t_2}$ (equations~\ref{eq:sws-42.a}a and~\ref{eq:sws-42.a}b) in the cross-section plane, singularities $\bm{n}^{t_3}$ and $\bm{n}^{t_4}$ (equations~\ref{eq:sws-42.a}c and~\ref{eq:sws-42.a}d) are located in the vertical plane orthogonal to the cross-section. Similar to Figure~\ref{fig:sws-08.1}d, the cyan patches in (b) are the planar bases $\bm{g}^Q$ of internal refraction cones formed at triple singularities. 
	} \vspace{+2mm}
	\label{fig:sws-09.1}
\end{figure}

Constructing an orthorhombic model for which the slowness components~\ref{eq:sws-42.7} are real-valued is not difficult. For example, a solid described by the stiffness matrix
\begin{equation} \label{eq:sws-43.9}
\def\arraystretch{1.0}
\bm{c} = \left( \begin{array}{c c r c c c}
7 & -3 & -2 & 0 & 0 & 0 \\
~ & ~~8 & -1 & 0 & 0 & 0 \\
~ &  ~ &  9 & 0 & 0 & 0 \\
~ & \mathrm{SYM} & ~ &  1 & 0 & 0 \\
~ &  ~ &  ~ &           ~ & 2 & 0 \\
~ &  ~ &  ~ &           ~ & ~ & 3 \\
\end{array} \right) \! ,
\def\arraystretch{1.0}
\end{equation}
satisfying conditions~\ref{eq:sws-42.6}, has four distinct triple singularities 
\begin{subnumcases}{\label{eq:sws-42.a}}
\bm{n}^{t_1} = [+1, \, +1, \, +1]/\sqrt{3} \, , & \\			
\bm{n}^{t_2} = [+1, \, +1, \, -1]/\sqrt{3} \, , & \\	
\bm{n}^{t_3} = [+1, \, -1, \, +1]/\sqrt{3} \, , & \\
\bm{n}^{t_4} = [-1, \, +1, \, +1]/\sqrt{3} \, , & 
\end{subnumcases}
directions of the two of them, $\bm{n}^{t_1}$ and $\bm{n}^{t_2}$, marked by the cyan arrows in Figure~\ref{fig:sws-09.1}a. 

The P-wave sheet of the group-velocity surface in Figure~\ref{fig:sws-09.1}b displays another remarkable shape~| two smooth square pyramids, connected together at their horizontal bases and symmetric with respect to the horizontal, as they should be to maintain the central symmetry of group-velocity surfaces. And like Figure~\ref{fig:sws-08.1}d, the faces of the pyramids exhibit large planar patches, colored in cyan.

\section{Vertical transverse isotropy} \label{ch:sws-triple-VTI} 

The transition from triple singularities at the vertical in orthorhombic media to those in vertically transversely isotropic (VTI) media is straightforward. Because $c_{13} = c_{23}$ for vertical transverse isotropy, the internal refraction cone~\ref{eq:sws-39} is circular,
\begin{equation}\label{eq:sws-43}
\def\arraystretch{2.2}
\bm{g}^t = \bm{g}(\bm{n}^t, \, \varphi_1, \, \varphi_2) = \left[ \begin{array}{c}
\dfrac{c_{13} + c_{55}}{2 \, V_t} \, \sin 2 \, \varphi_1 \, \cos \varphi_2  \\
\dfrac{c_{13} + c_{55}}{2 \, V_t} \, \sin 2 \, \varphi_1 \, \sin \varphi_2  \\
V_t
\end{array} \right] \! ;
\def\arraystretch{1.0}
\end{equation}
its horizontal base $\bm{g}^Q$ has a radius of 
\begin{equation} \label{eq:sws-43.1}
\max \left( g_1^t \right) = \max \left( g_2^t \right) = \dfrac{1}{2} \left(\dfrac{c_{13}}{V_t} + V_t \right) \! .
\end{equation}

\begin{figure}
	\centering
	\begin{tabular}{c c}
		~ \qquad (a) & ~ \quad (b) \\
		\includegraphics[height=0.34\textwidth]{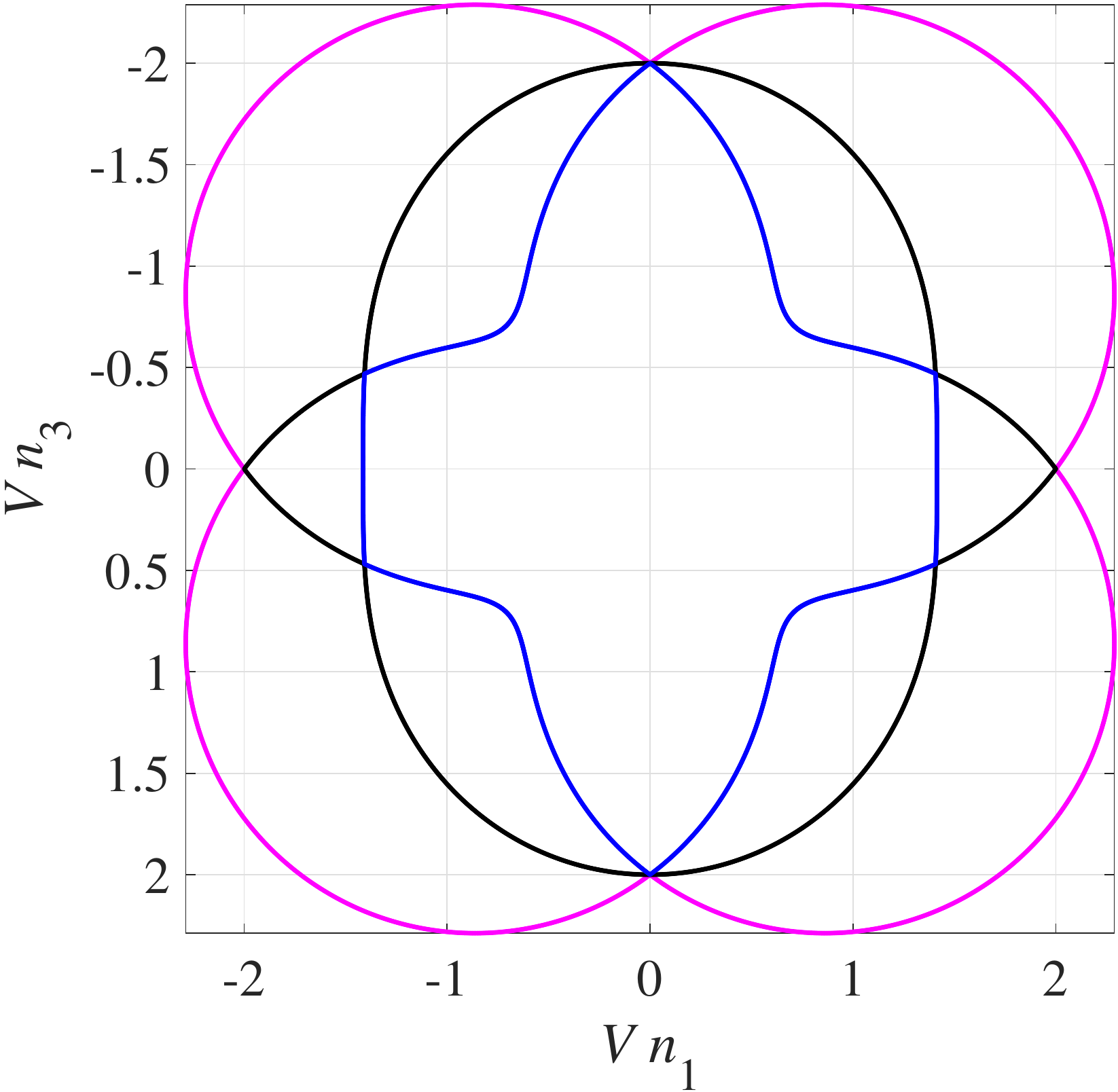} &
		\includegraphics[height=0.34\textwidth]{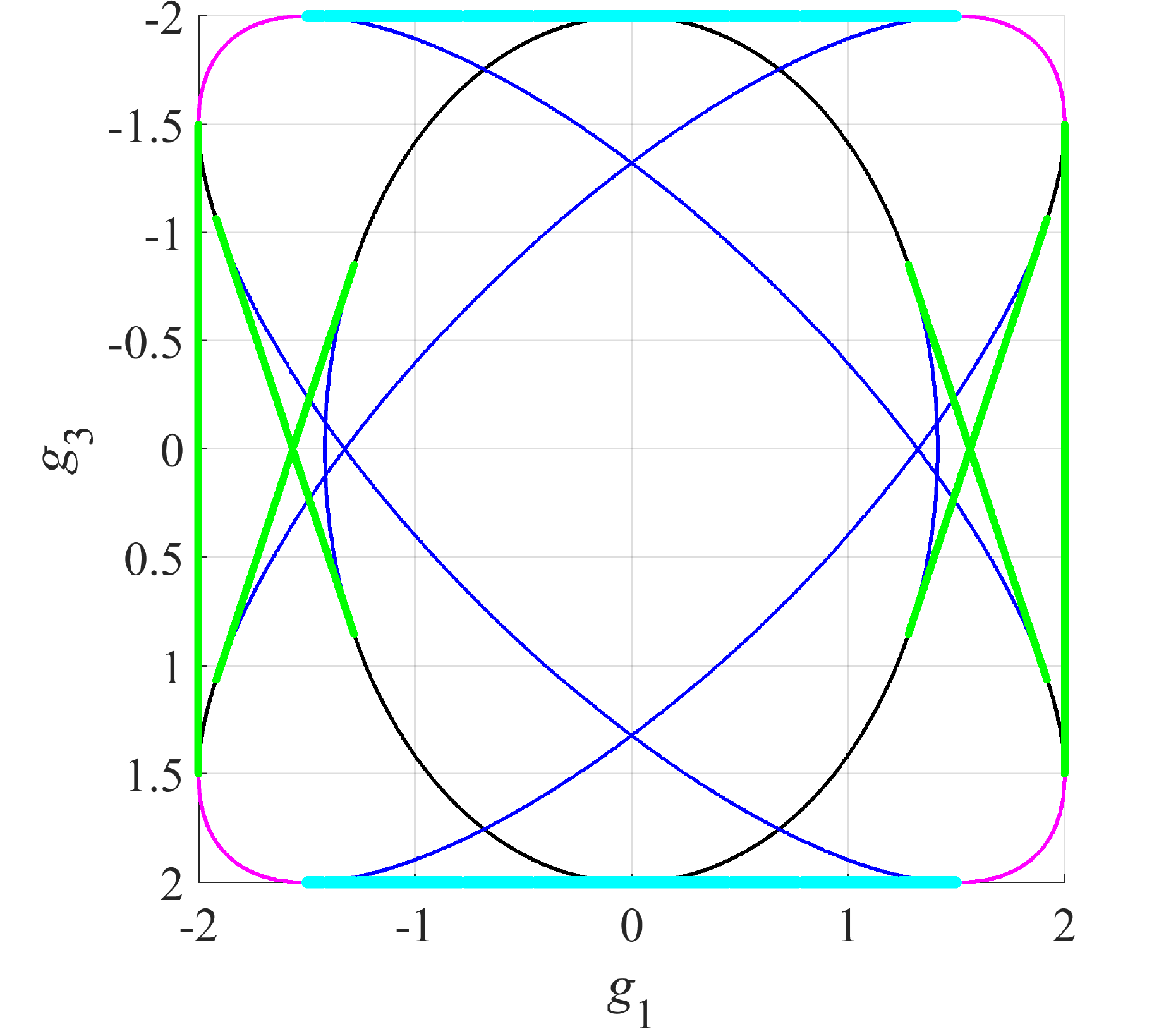} \\
	\end{tabular}
	\caption{(a) Phase- and (b) group-velocity surfaces of the P- (magenta), S$_1$- (black), and S$_2$-waves (blue) in the $[\bm{x}_1, \, \bm{x}_3]$ plane of VTI model~\ref{eq:sws-45}. The bases $\bm{g}^E$ of degenerative internal refraction cones corresponding to the intersection singularities  are displayed in green, the bases $\bm{g}^Q$ of internal refraction cones corresponding to the triple singularities at the vertical~| in cyan. 
	} \vspace{-2mm}
	\label{fig:sws-09}
\end{figure}

Like triple and conical singularities in triclinic and orthorhombic solids, triple and intersection singularities can coexist in VTI media, as exemplified in Figure~\ref{fig:sws-09} computed for a model that has the stiffness matrix
\begin{equation} \label{eq:sws-45}
\def\arraystretch{1.0}
\bm{c} = \left( \begin{array}{c c c c c c}
4 &  0 &  2 & 0 & 0 & 0 \\
~ &  4 &  2 & 0 & 0 & 0 \\
~ &  ~ &   4 & 0 & 0 & 0 \\
~ & \mathrm{SYM} & ~ &  4 & 0 & 0 \\
~ &  ~ &  ~ &           ~ & 4 & 0 \\
~ &  ~ &  ~ &           ~ & ~ & 2 \\
\end{array} \right) \! .
\def\arraystretch{1.0}
\end{equation}
A notable feature of Figure~\ref{fig:sws-09}b is an atypically small portion of the purely P-wave branch (magenta) of the total group-velocity surface, also observed in Figures~\ref{fig:sws-08.1}b (magenta) and~\ref{fig:sws-08.1}d (copper).

\section{Discussion} \label{ch:sws-discussion} 

Triple singularities, admittedly unusual objects, entailing the presence of finite-size planar patches on group-velocity surfaces, raise a valid question about practical importance of their mathematical possibility. Because the existence of triple singularities rests upon equality-type constraints (equations~\ref{eq:sws-31}) that can be destroyed by an arbitrary triclinic perturbation of the stiffness tensor,  \citet{Alshitsetal1985} suggested that ``triple [singularities] do not occur.'' The logic behind this statement is fascinating, linking it to the philosophical in some sense subject of the presence of anisotropic symmetries in rocks as opposed to those in crystals, where symmetry of a crystal inherits the exact symmetry of its atomic lattice. Because symmetry of a rock sample can be only approximate, and the concept of approximate symmetry is barely touched in the literature, it seems appropriate to briefly discuss it.

Elastic properties and symmetry of a volume of core or geologic formation can be computed given the precise knowledge of microstructure of the volume and the stiffness tensors $\bm{c}(\bm{x})$ at its physical points $\bm{x}$. The process of computing the overall or effective elastic properties, known as homogenization, has been covered in several books\cite{NematNasser1999, Milton2002, Kachanov2003Handbook} and applied to fractured solids\cite{GrechkaKachanov2006, GrechkaKachanovXfrac2006, GrechkaetalSfrac2006, Grechka2007IJF, Grechka2007}. Because rocks are composed of diverse anisotropic minerals that are not perfectly aligned and contain dry or fluid-filled pores and fractures, the overall symmetry of rocks usually comes out triclinic, as confirmed by suitable seismic data\cite{DewanganGrechka2003, GrechkaYaskevich2014} and numerical experiments. Hence, the stiffness matrixes of solids with symmetries lower than triclinic should be deemed as mere approximations of a more complex reality. Yet, explicitly imposing equality-type constrains on the stiffness tensor to make it symmetric~| orthorhombic, transversely isotropic, or even isotropic, as expressed by the zeros and coinciding elements in the corresponding stiffness matrixes~| has proven extremely useful in numerous rock physics, seismic, and seismological applications. Therefore, the fact that an arbitrary triclinic perturbation breaks down elastic symmetries is irrelevant; and as long as an adopted symmetry, \emph{understood as an approximation}, happens to be helpful in solving particular problems, its inexactness is not only acceptable but even desirable for simplifying the ensuing computations and conclusions.

Likewise, slight violation of constraints~\ref{eq:sws-31}, while formally destroying the triple singularity as a mathematical entity, would not alter seismic signatures in any significant way; and because P-to-S$_1$ singularities occur in certain types of woods\cite{Musgrave1970}, the existence of materials exhibiting some semblance of triple singularities might be possible.

\section{Conclusions}

The following features of triple singularities have been established.
\begin{itemize} \vspace{-2mm}
	\item The group-velocity vectors at a triple singularity fill solid cones. \vspace{-2mm}
	
	\item These internal refraction cones are generally quartic, although they can degenerate to quadratic~| elliptic or circular~| in more symmetric solids than triclinic. \vspace{-2mm}
	
	\item The base of an internal refraction cone is planar, giving rise to planar patches shared by group-velocity surfaces of the P-, S$_1$-, and S$_1$-waves. \vspace{-2mm}
	
	\item The maximum number of triple singularities is equal to 4, smaller than that of conventional (double) singularities, known to be 16. \vspace{-2mm}
	
	\item If multiple triple singularities are present, the phase velocities along all of them are exactly equal.
\end{itemize} \vspace{-2mm}

\section{Acknowledgments}

I would like to thank Yuriy Ivanov for spotting typos in equations~\ref{eq:sws-35} and~\ref{eq:sws-36}. The typos are now corrected.



\bibliography{refs-vg-DCP}  

\end{document}